\pdfoutput=1
\documentclass[twocolumn]{aastex631}
\graphicspath{{./}{figures/}}
\usepackage[T1]{fontenc}

\usepackage{soul}
\usepackage{amsmath}

\shorttitle{Cloud-to-Core Gas Flows in Mon~R2~HFS}
\shortauthors{L.~K. Dewangan et al.}

\begin{document}
\title{Hierarchical Hub Formation in Mon~R2: Evidence for Cloud-to-core Gas Flows}

\author[0000-0001-6725-0483]{L.~K. Dewangan}
\affiliation{Astronomy \& Astrophysics Division, Physical Research Laboratory, Navrangpura, Ahmedabad 380009, India.}
\email{lokeshd@prl.res.in}

\author[0009-0001-2896-1896]{O.~R. Jadhav}
\affiliation{Astronomy \& Astrophysics Division, Physical Research Laboratory, Navrangpura, Ahmedabad 380009, India.}
\affiliation{Indian Institute of Technology Gandhinagar Palaj, Gandhinagar 382355, India.}

\author[0000-0002-7367-9355]{A.~K. Maity}
\affiliation{Faculty of Engineering, Gifu University, 1-1 Yanagido, Gifu 501-1193, Japan.}

\author[0000-0001-5731-3057]{Saurabh Sharma}
\affiliation{Aryabhatta Research Institute of Observational Sciences, Manora Peak, Nainital 263002, India.}

\author[0000-0002-6740-7425]{Ram~K. Yadav}
\affiliation{National Astronomical Research Institute of Thailand (Public Organization), 260 Moo 4, T. Donkaew, A. Maerim, Chiangmai 50180, Thailand.}

\begin{abstract}
We present a multi-scale kinematic study of the hub-filament system in Mon R2 using JCMT $^{13}$CO(3--2) and ALMA C$^{18}$O(1--0) observations to investigate how molecular gas converges toward the central hub and connects across cloud-to-core scales. 
The $^{13}$CO data reveal several velocity-coherent filaments with diverse position angles converging toward the central region hosting massive stars and stellar clusters, where distinct velocity components become spatially superposed. Two dominant velocity regimes at $\sim$8.8--10.2 and $\sim$10.4--11.8 km s$^{-1}$ are identified, with their velocity separation decreasing toward the central hub. 
The position--position--velocity and longitude--velocity maps of the $^{13}$CO emission exhibit a butterfly/X-shaped morphology, consistent with converging motions. A phenomenological rotating-and-infalling envelope model reproduces the longitude-velocity morphology, indicating that rotation and infall may contribute to the observed kinematics, although the model does not uniquely identify the underlying mechanism.
At higher resolution, the ALMA C$^{18}$O observations resolve the molecular gas into multiple velocity-coherent filaments spanning $\sim$8--13 km s$^{-1}$, with distinct kinematic domains at $\sim$9--10.5, 10.5--11.5, and 12--13 km s$^{-1}$ converging toward the hub. The ALMA and JCMT position-velocity diagrams show consistent velocity structures across their respective spatial scales, with the ALMA-resolved features embedded within the broader molecular-gas velocity field traced by JCMT, demonstrating a multi-scale kinematic connection. These results support a hierarchical picture in which converging flows and filamentary accretion contribute to hub growth and ongoing star formation in Mon R2, while other dynamical processes may also contribute to the observed kinematics.
\end{abstract}
%
%
%
%
\section{Introduction} 
\label{sec:intro}
In recent years, hub-filament systems \citep[HFSs;][]{myers09} have emerged as fundamental structures associated with the formation of massive OB stars ($>$ 8 $M_{\odot}$) and stellar clusters. The converging filaments in these systems are thought to funnel material into dense hubs, promoting gas accumulation, fragmentation, and subsequent star formation \citep[e.g.,][]{Motte+2018,padoan20,semadeni09,zhou22,maity23,dewangan25n}. 
Nevertheless, the physical mechanisms responsible for the formation and growth of hubs remain uncertain. In particular, it is not yet clear whether hub assembly is driven by large-scale gravitational inflows, converging flows, cloud-cloud collisions, or other dynamical processes. This observational study focuses on the cluster-forming HFS in Monoceros R2 (Mon~R2; extent $\le$ 0.8 pc; \citealt{morales19,kumar22,dewangan25s}), one of the nearest ($\sim$830 pc; \citealt{racine68,herbst76}) and best-studied regions of massive star formation \citep{andersen06,beckwith76,choi00,carpenter08,didelon15,dierickx15,downes75,fuente10,
giannakopoulou97,henning92,massi85,meyers91,tafalla97}. 
Owing to its proximity, Mon~R2 provides a unique opportunity to investigate the physical processes governing the assembly and evolution of dense hubs and the formation of massive stars within their natal environments.

Using IRAM $^{13}$CO and C$^{18}$O observations, \citet{morales19} established the presence of an HFS in Mon~R2 and identified spiral-like gas motions toward the central region. 
They reported that the hub kinematics reveal rotation and infall associated with gas convergence toward the stellar cluster, alongside the expansion of the central ultra-compact H\,{\sc ii} region \citep{didelon15} into the surrounding envelope. More recently, \citet{dewangan25s}, using Atacama Large Millimeter/submillimeter Array (ALMA) observations along with complementary multi-wavelength data sets, identified a molecular ring encircling an infrared ring, with several molecular filaments appearing to converge toward the molecular ring (see Figure~11 in their paper). The central region of the Mon~R2~HFS harbors an embedded cluster of young stellar objects (YSOs) along with several massive stars. They suggested that Mon~R2~HFS evolved from an infrared-quiet to an infrared-bright phase through the interplay of gas accretion and massive-star feedback. At least five infrared sources (IRS~1--5) were reported toward the molecular ring \citep[see Figure~1a in][]{dewangan25s}. Previous studies have examined the Mon~R2~HFS over a wide range of spatial scales, revealing its hierarchical filamentary structure and an active star-forming hub \citep[e.g.,][]{morales19,kumar22,dewangan25s}. 
However, the kinematic connection between the cloud-scale molecular gas and the dense central hub remains poorly constrained, leaving the origin and assembly history of the hub uncertain. 

In this paper, we studied the gas dynamics within the Mon~R2~HFS using the ALMA C$^{18}$O(1--0) line data and the James Clerk Maxwell Telescope (JCMT) $^{13}$CO(3--2) data cube of Mon~R2. By combining these complementary data sets, we trace gas kinematics over a wide range of spatial scales, from parsec-scale molecular clouds to sub-parsec hub environments. Our analysis reveals, for the first time, a multi-scale kinematic connection between large-scale gas convergence and filamentary accretion onto core-scale regions in Mon~R2~HFS.

The structure of this paper is as follows. Section~\ref{sec:obser} presents the observational data sets employed in this study. The observational results are described in Section~\ref{sec:xxdata}. In Section~\ref{sec:disc}, we discuss the implications of these results for Mon~R2. Finally, Section~\ref{sec:sumc} summarizes our main conclusions. 
\section{Data sets}
\label{sec:obser}
\begin{figure*}
\center
\includegraphics[width=\textwidth]{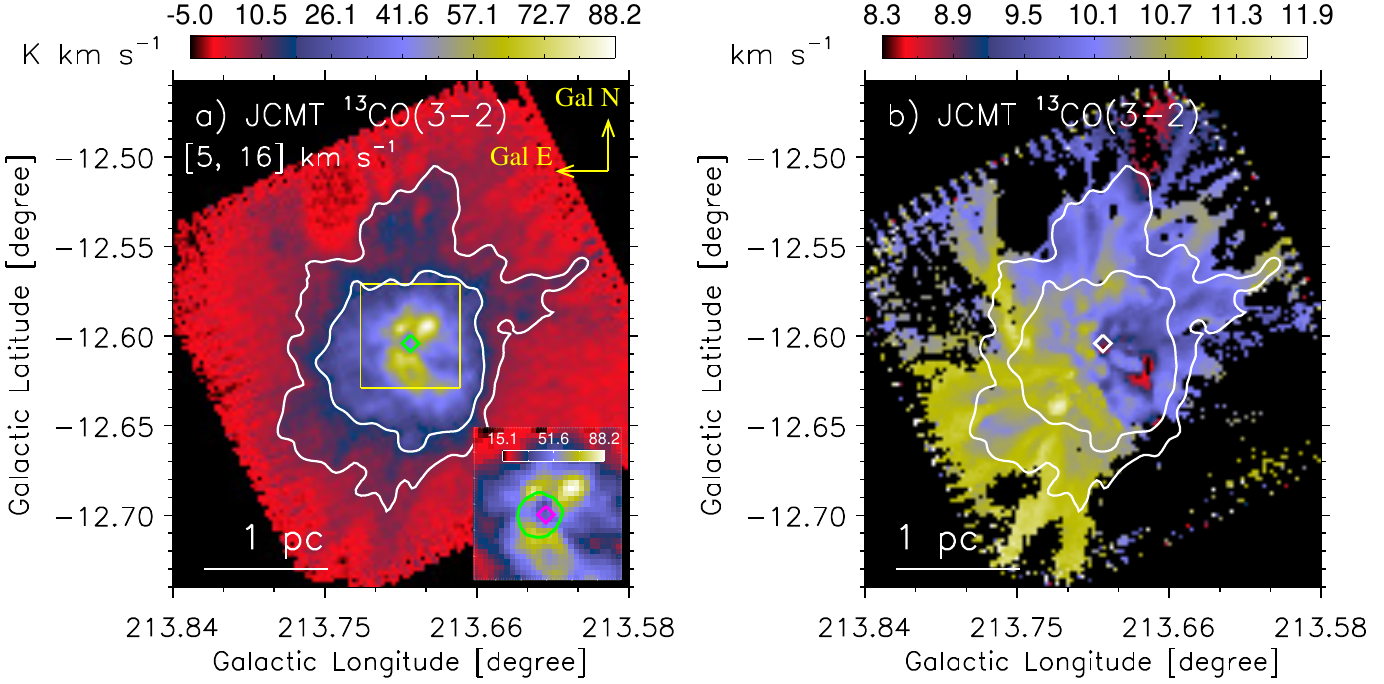}
\caption{a) JCMT $^{13}$CO(3--2) moment-0 map integrated over the velocity range 5--16 km s$^{-1}$. The yellow box outlines the area investigated using the ALMA C$^{18}$O(1--0) observations presented in this work. White contours show the $^{13}$CO(3--2) integrated intensity at 8.8 and 17.5 K km s$^{-1}$. The inset shows an enlarged view of the boxed region in the $^{13}$CO(3--2) moment-0 map, overlaid with the NVAS 1.49 GHz radio continuum emission shown by the green contour at 16.4 mJy beam$^{-1}$. b) Intensity-weighted mean velocity (moment-1) map of the JCMT $^{13}$CO(3--2) emission, with the same integrated-intensity contours as in Figure~\ref{fig1}a. 
The scale bar corresponds to 1 pc at the adopted distance of 830 pc. The position of IRS~2 is marked by a diamond in both main panels and the inset.}
\label{fig1}
\end{figure*}

We used the ALMA Band~3 C$^{18}$O($J$ = 1--0) line data ($\nu_{\rm rest}=109.7822$~GHz) obtained from project \#2016.1.01144.S (PI: Trevi{\~n}o-Morales, Sandra). The observations were carried out during ALMA Cycle~4 using the 12m array in the C40-1 configuration. The synthesized beam of the C$^{18}$O(1--0) data is $\sim$3\rlap.{$''$}68 $\times$ 2\rlap.{$''$}64 (or $\sim$3055 $\times$2190 AU; pixel scale $\sim$0\rlap.{$''$}52), and the corresponding brightness sensitivity is $\sim$30~mJy~beam$^{-1}$ or $\sim$0.31~K. To investigate the larger-scale molecular gas distribution and kinematics, we utilized the science-ready JCMT $^{13}$CO(3--2) spectral data cube ($\nu_{\rm rest}$ $\sim$330.587960 GHz; proposal ID = R19BP001), retrieved from the JCMT Science Archive hosted by the Canadian Astronomy Data Centre (CADC). The JCMT observations were conducted using the Heterodyne Array Receiver Programme coupled with the Auto-Correlation Spectral Imaging System \citep[HARP/ACSIS;][]{buckle09}. The resulting line cube has a pixel scale of $\sim$7\rlap.{$''$}3 and an angular resolution of $\sim$14$''$ or 11620 AU. The noise per channel is about 0.7~K on the T$_{A}$ scale.  
The JCMT $^{13}$CO(3--2) and ALMA C$^{18}$O(1--0) observations have velocity resolutions of 0.055 and 0.083 km s$^{-1}$, respectively. 
In addition, archival radio continuum data at 1.49 GHz (beam size $\sim$15$''$, rms =167 $\mu$Jy) were obtained from the NRAO VLA Archive Survey \citep[NVAS;][]{crossley07} to examine the ionized gas emission associated with the target region. The direction in each Figure~\ref{fig1} of this paper is shown in the Galactic coordinates. 

\begin{figure*}
\center
\includegraphics[width=\textwidth]{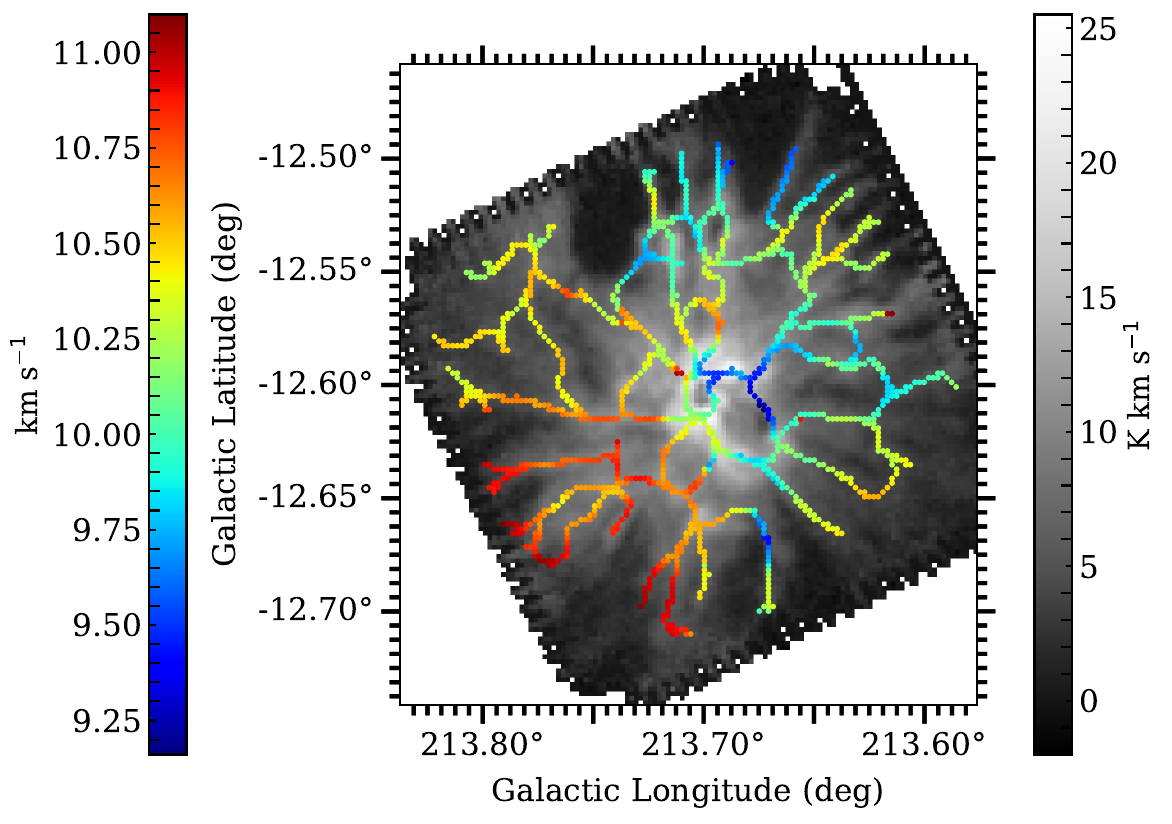}
\caption{The JCMT $^{13}$CO(3--2) moment-0 map showing the filamentary structures identified using the DisPerSE algorithm \citep{sousbie11}. The colored curves represent the skeletons of the identified filaments, with the color indicating the corresponding $^{13}$CO velocity along each filament, as shown by the left color bar. The grayscale background represents the  $^{13}$CO integrated intensity, with the intensity scale shown on the right. The coherent velocity variations along the filamentary skeletons highlight their kinematic structure.}
\label{xfig1}
\end{figure*}
\begin{figure*}
\center
\includegraphics[width=\textwidth]{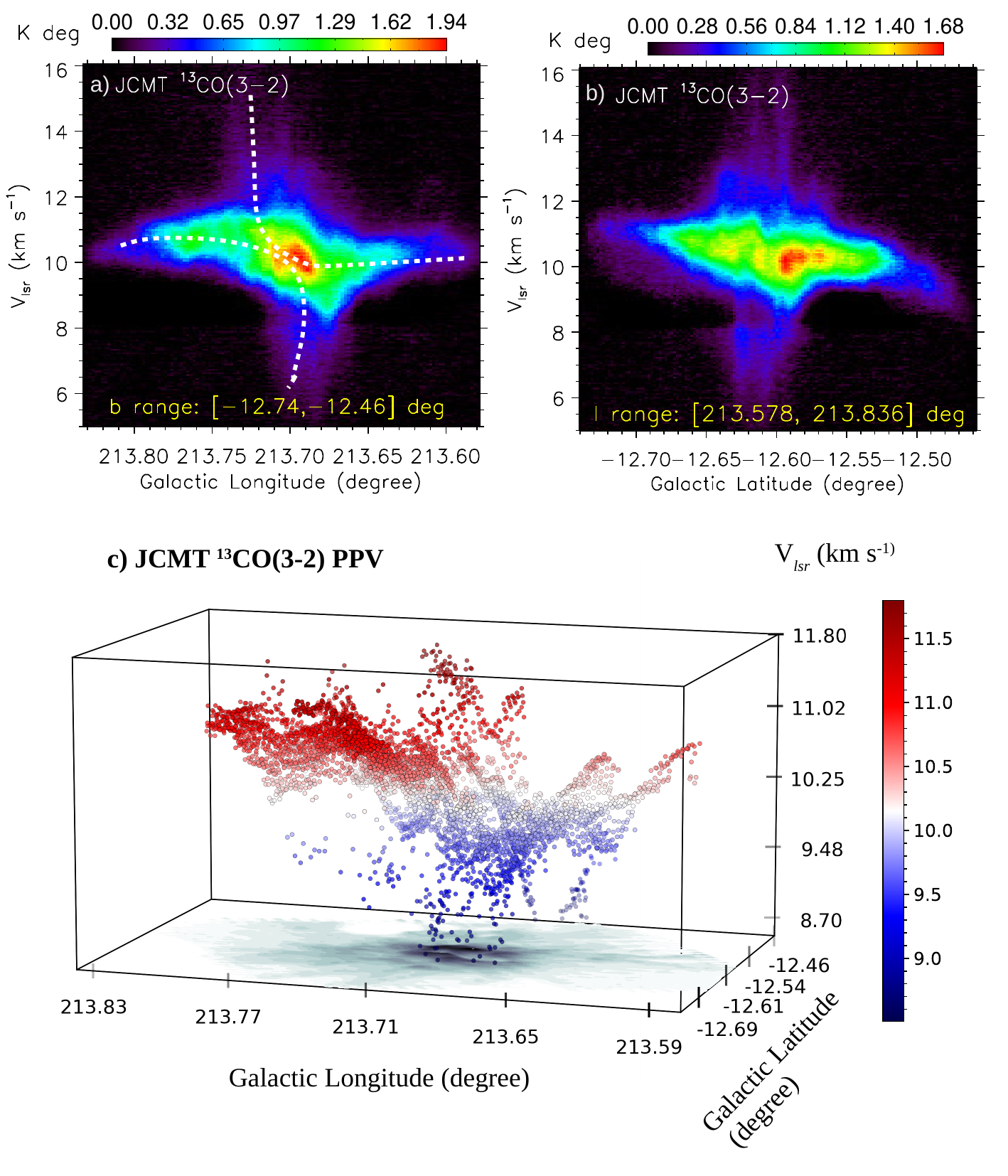}
\caption{a) Longitude-velocity diagram of the JCMT $^{13}$CO(3--2) emission. 
The dotted curves outline the prominent butterfly/X-shaped morphology, and the latitude range adopted for the integration is indicated. b) Latitude-velocity diagram of the $^{13}$CO(3--2) emission, with the longitude range used for the integration indicated. c) PPV map of the $^{13}$CO(3--2) emission.} 
\label{fig2}
\end{figure*}
\section{Results}
\label{sec:xxdata}
\subsection{Gas kinematics at cloud scales}
\label{subsec:cloud}
To study the large-scale molecular gas structure in Mon~R2, we analyzed unpublished JCMT $^{13}$CO(3--2) observations in the velocity interval of [5, 16] km s$^{-1}$. Figure~\ref{fig1}a presents the corresponding integrated-intensity (moment-0) map, which reveals an extended molecular gas distribution encompassing the Mon~R2 HFS. 
The inset highlights the central boxed region, showing the $^{13}$CO(3--2) emission together with the NVAS 1.49 GHz radio continuum contour. 
The radio continuum emission traces the H\,{\sc ii} region in Mon~R2.  
Figure~\ref{fig1}b shows the corresponding moment-1 map. The velocity field reveals multiple elongated structures associated with distinct velocity components. 
The southeastern region is predominantly redshifted, whereas the northwestern region is mainly blueshifted. However, the velocity field is highly structured, exhibiting multiple velocity components and localized velocity variations rather than a simple, monotonic gradient. 
The transition between the two velocity regimes occurs near the cloud center. Figure~\ref{afg2} presents the channel maps of the JCMT $^{13}$CO(3--2) emission (see also Appendix~\ref{subsec:a1}). The molecular-gas morphology varies systematically with velocity, with the low- and high-velocity emission exhibiting distinct spatial distributions. Toward the central region, these velocity components become spatially superposed, coincident with the intersection of several filamentary structures.  

To further characterize the elongated structures and examine their spatial and kinematic connection to the central hub, we applied the DisPerSE algorithm \citep{sousbie11} to the JCMT $^{13}$CO(3--2) moment-0 map. DisPerSE is a publicly available tool that identifies topological structures in a density field by locating critical points and connecting them within the framework of Morse theory. The resulting filament skeletons are shown in Figure~\ref{xfig1}, overlaid on the grayscale $^{13}$CO integrated-intensity map, with the corresponding intensity scale indicated by the right color bar. To investigate the velocity structure along these filaments, we extracted $^{13}$CO spectra within a 14$''$ aperture, corresponding to the JCMT beam size, centered on successive positions along each skeleton. The spectra were fitted with Gaussian components to determine the corresponding centroid velocities. Where two Gaussian components were required to adequately reproduce the observed profile, the mean of their centroid velocities was adopted as the representative velocity at that position. Since the spectra cover a relatively narrow velocity range, the separation between the two fitted components is limited to $\lesssim 1$ km s$^{-1}$, comparable to the characteristic velocity dispersion of the region. The resulting centroid velocities are encoded along the filament skeletons, with the color scale indicating the $^{13}$CO velocity and the left color bar showing the corresponding velocity range.
 
Several velocity-coherent filaments, having a broad range of position angles, converge toward the central region, where distinct velocity components overlap. The filament skeletons therefore reveal a spatially connected network of molecular structures with different characteristic velocities that intersect near the Mon~R2 hub. The velocity-coded skeletons further show that the velocity variations are not randomly distributed but are associated with individual filamentary structures and their convergence toward the central region. These observations demonstrate that the molecular gas surrounding the hub is organized into multiple velocity-coherent components that become spatially superposed near the central star-forming region.

Given the complex orientation of the filamentary network around the central hub in Mon~R2 (Figure~\ref{xfig1}), a single position-velocity (PV) cut along a specific filament would not adequately represent the large-scale velocity structure. We therefore constructed complementary longitude-velocity and latitude-velocity diagrams by integrating the JCMT $^{13}$CO(3--2) emission over the corresponding spatial coordinates. This approach allows the velocity structure to be examined across the full molecular-gas distribution without assuming that the velocity gradient is aligned with either the longitude or latitude direction. Figure~\ref{fig2}a presents the longitude-velocity diagram obtained by integrating the emission over the latitude range of [$-$12.74$^{\circ}$, $-$12.46$^{\circ}$]. The diagram displays a characteristic butterfly- or X-shaped morphology centered near the systemic velocity, $V_{\rm sys}\sim$10.25 km s$^{-1}$ (see dotted curves in Figure~\ref{fig2}a). The emission is distributed over a broad velocity range toward the central longitudes, whereas the velocity extent becomes narrower toward the outer parts of the cloud. The redshifted and blueshifted emission also exhibit a substantial degree of spatial overlap near the central longitude. Figure~\ref{fig2}b shows the corresponding latitude-velocity diagram, obtained by integrating over the longitude range of [213.578$^{\circ}$, 213.836$^{\circ}$]. A similar butterfly/X-shaped distribution is evident in this projection, with the broadest velocity extent occurring around the central latitude. 

The consistency between the longitude-velocity and latitude-velocity diagrams demonstrates that the complex velocity structure is present along both spatial directions. 
In both projections, distinct red- and blueshifted emission components are observed on either side of $V_{\rm sys}$,  together with intermediate-velocity emission near the central region. The resulting pattern differs from a simple monotonic velocity gradient across the cloud, as would be expected for an idealized large-scale rotating structure. In an idealized rotating structure, the redshifted and blueshifted emission are expected to exhibit an organized spatial separation associated with the rotation axis. The observed longitude-velocity and latitude-velocity diagrams, however, show a more complex butterfly/X-shaped velocity distribution, with multiple velocity components and substantial spatial overlap toward the central region. Emission spanning the velocity interval between the two principal velocity regimes is also present near the center. These characteristics indicate that the observed velocity field cannot be adequately described by a simple monotonic velocity gradient alone. We therefore investigate the contribution of rotation together with infall using a beam- and spectrally convolved kinematic model in Appendix~\ref{subsec:a2} (see also Figure~\ref{nnafg3}).

Figure~\ref{fig2}c presents the position-position-velocity (PPV) distribution of the JCMT $^{13}$CO(3--2) emission over the region shown in Figure~\ref{fig1}a. Although the viewing angle is similar to that in Figure~\ref{fig2}a, the PPV representation retains the two-dimensional spatial information together with the velocity dimension, allowing the spatial distribution and continuity of the different velocity components to be examined simultaneously. It therefore provides a complementary representation to the longitude-velocity and latitude-velocity diagrams, particularly for relating the identified velocity components and intermediate-velocity emission to their locations within the filamentary structures. The PPV map was produced using SCOUSEPY \citep[][]{henshaw16,henshaw19,dewangan25s}, a tool designed to decompose complex spectra into individual velocity components. Two dominant velocity regimes can be identified: a redshifted component at $\sim$10.4--11.8 km s$^{-1}$ and a blueshifted component at $\sim$8.8--10.2 km s$^{-1}$. Both components are spatially extended and appear to approach one another toward the central hub located near $\mathrm{\it l}\approx213.68^{\circ}$. Emission spanning the velocity interval between these two regimes is also detected toward the central region. In PPV space, the molecular-gas distribution exhibits a V-shaped to butterfly-like morphology, with the velocity separation between the two principal components decreasing toward the central hub. Several filamentary branches extend toward the central region from different directions, where the distinct velocity components become increasingly spatially superposed. These features demonstrate that the molecular gas around the Mon~R2 hub comprises multiple spatially extended and kinematically distinct components with a systematic change in their spatial and velocity relationship toward the central region.
\begin{figure*}
\center
\includegraphics[width=\textwidth]{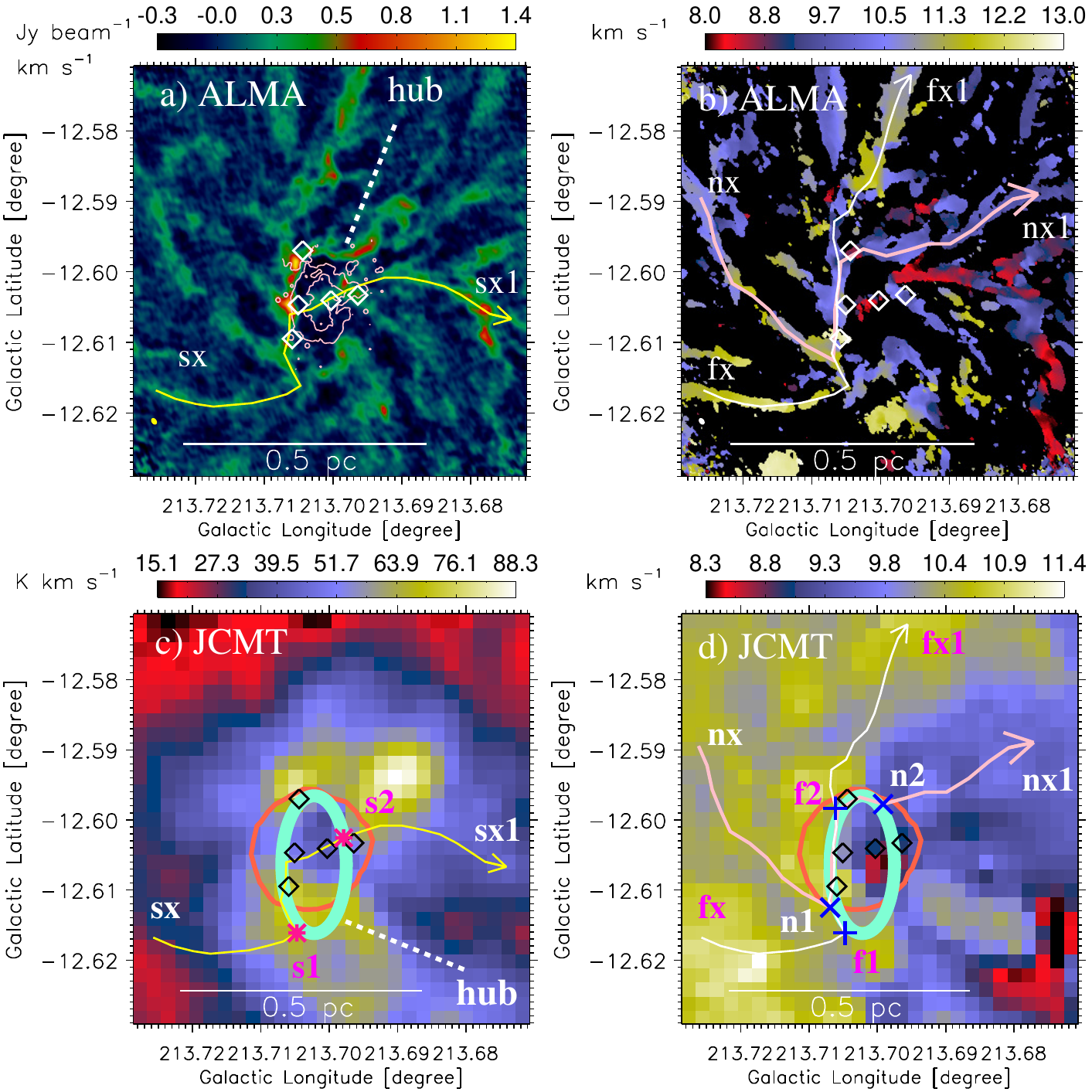}
\caption{Molecular gas morphology and velocity structure toward the central region of the Mon R2 hub. a) ALMA C$^{18}$O(1--0) moment-0 map of the region enclosed by the yellow box in Figure~\ref{fig1}a. The infrared ring is shown by pink contours, and the path sx--sx1 used to extract the PV diagram is indicated. 
b) ALMA C$^{18}$O(1--0) moment-1 map. Two different paths fx--fx1 and nx--nx1 adopted for the PV analysis are marked. c) JCMT $^{13}$CO(3--2) moment-0 map covering the same region as in Figure~\ref{fig3}a (see the yellow box in Figure~\ref{fig1}a). The NVAS 1.49 GHz radio continuum emission is overlaid as a tomato contour at 16.4 mJy beam$^{-1}$. The thick aquamarine ellipse marks the molecular ring/hub, while the asterisks mark the locations where the sx--sx1 path connects to the hub, labeled s1 and s2. d) JCMT $^{13}$CO(3--2) moment-1 map corresponding to the panel ``c''. The ``+'' and ``$\times$'' symbols mark the connections of the fx--fx1 and nx--nx1 paths with the hub, labeled f1, f2 and n1, n2, respectively. In all panels, diamonds mark the positions of five infrared sources. The scale bar corresponds to 0.5 pc. The filled ellipse in panels ``a'' and ``b'' represents the beam size of the ALMA C$^{18}$O(1--0) observations.} 
\label{fig3}
\end{figure*}
\subsection{Core-scale gas kinematics and its connection to cloud-scale kinematics}
\label{subsec:core}
\subsubsection{ALMA and JCMT moment maps}
\label{1subs}
To investigate the gas kinematics at core scales, we examined the ALMA C$^{18}$O(1--0) line data over a velocity range of $\sim$8--13 km s$^{-1}$ toward the central region of Mon~R2 (see the yellow box in Figure~\ref{fig1}a).  
Figures~\ref{fig3}a and \ref{fig3}b present the corresponding moment-0 and moment-1 maps, respectively, which were previously analyzed by \citet{dewangan25s}. 
These maps are reproduced here to facilitate comparison with the new kinematic analysis reported in this study. Within the spatial scale of 0.8 pc, the C$^{18}$O moment-0 map reveals at least 6--8 prominent filamentary arms converging toward a compact central hub/molecular ring, which surrounds the 
infrared ring (see pink contours in Figure~\ref{fig3}a). We have also selected three paths (fx--fx1, sx--sx1, and nx--nx1; see Figures~\ref{fig3}a and~\ref{fig3}b) that follow the most prominent velocity-coherent filamentary structures seen in the ALMA moment maps and are used to extract the corresponding PV diagrams.

In Figure~\ref{fig3}b, the ALMA moment-1 map reveals a complex velocity field characterized by several distinct velocity regimes ($\sim$9--10.5, 10.5--11.5, and 12--13 km s$^{-1}$) that converge toward the central hub/molecular ring. This ring-like structure seems to be the main gravitational potential well of the system. 
Near the hub, the velocity distribution becomes increasingly fragmented and irregular. Channel maps of the ALMA C$^{18}$O(1--0) emission demonstrate that these filaments are velocity-coherent dynamical structures, indicating that they represent physically connected gas flows (see Figure~\ref{afg3} and also Appendix~\ref{subsec:a1x}). The velocity variation across the central $\sim$0.1 pc region is approximately $\Delta V$ $\sim$2--3 km s$^{-1}$. Rather than tracing a single ordered gradient, the observed kinematics are consistent with the convergence of multiple gas flows toward the hub. 

We have also generated the moment-0 and moment-1 maps of the JCMT $^{13}$CO(3--2) emission over the same region covered by the ALMA observations (see Figure~\ref{fig3}a). The resulting JCMT $^{13}$CO(3--2) moment-0 and moment-1 maps are presented in Figures~\ref{fig3}c and \ref{fig3}d, respectively, with a contour representing the radio continuum emission. The molecular ring/central hub is indicated by the aquamarine ellipse in these JCMT maps. Furthermore, the path sx--sx1 is shown in Figure~\ref{fig3}c (see also Figure~\ref{fig3}a), while the other two paths (fx--fx1 and nx--nx1) are also highlighted in Figure~\ref{fig3}d (see also Figure~\ref{fig3}b). The selected paths sx--sx1, fx--fx1, and nx--nx1 intersect the central hub at positions s1--s2, f1--f2, and n1--n2, respectively (see Figures~\ref{fig3}c and \ref{fig3}d), which serve as reference points for the subsequent PV analysis. Along these paths, the prominent filaments fx--f1, sx--s1, and nx--n1 are located on one side of the hub, while their corresponding filamentary extensions, f2--fx1, s2--sx1, and n2--nx1, extend on the opposite side. The segments f1--f2, s1--s2, and n1--n2 correspond to the portions of the selected paths that traverse the central hub/molecular-ring region.

The JCMT $^{13}$CO(3--2) moment-0 map shows extended molecular emission surrounding the compact C$^{18}$O structure traced by ALMA, while the corresponding moment-1 map reveals a coherent velocity pattern across the central hub and its surrounding molecular environment. A comparison between the ALMA C$^{18}$O and JCMT $^{13}$CO(3--2) emission indicates that the observed molecular-gas kinematics extend across multiple spatial scales.
This comparison does not imply a direct one-to-one correspondence between individual velocity components. In particular, the ALMA C$^{18}$O moment-1 map traces systematic velocity variations within the compact central region, whereas the lower-resolution JCMT $^{13}$CO(3--2) data reveal a corresponding velocity pattern over the more extended molecular environment. Despite probing different spatial scales and gas densities, the broadly consistent velocity patterns suggest a kinematic connection between the compact central gas traced by ALMA and the surrounding large-scale molecular material, rather than a velocity structure confined solely to the central region (see also Figures~\ref{afg2} and~\ref{afg3}). 
\subsubsection{ALMA and JCMT position-velocity diagrams}
\label{2subs}
\begin{figure*}
\center
\includegraphics[width=\textwidth]{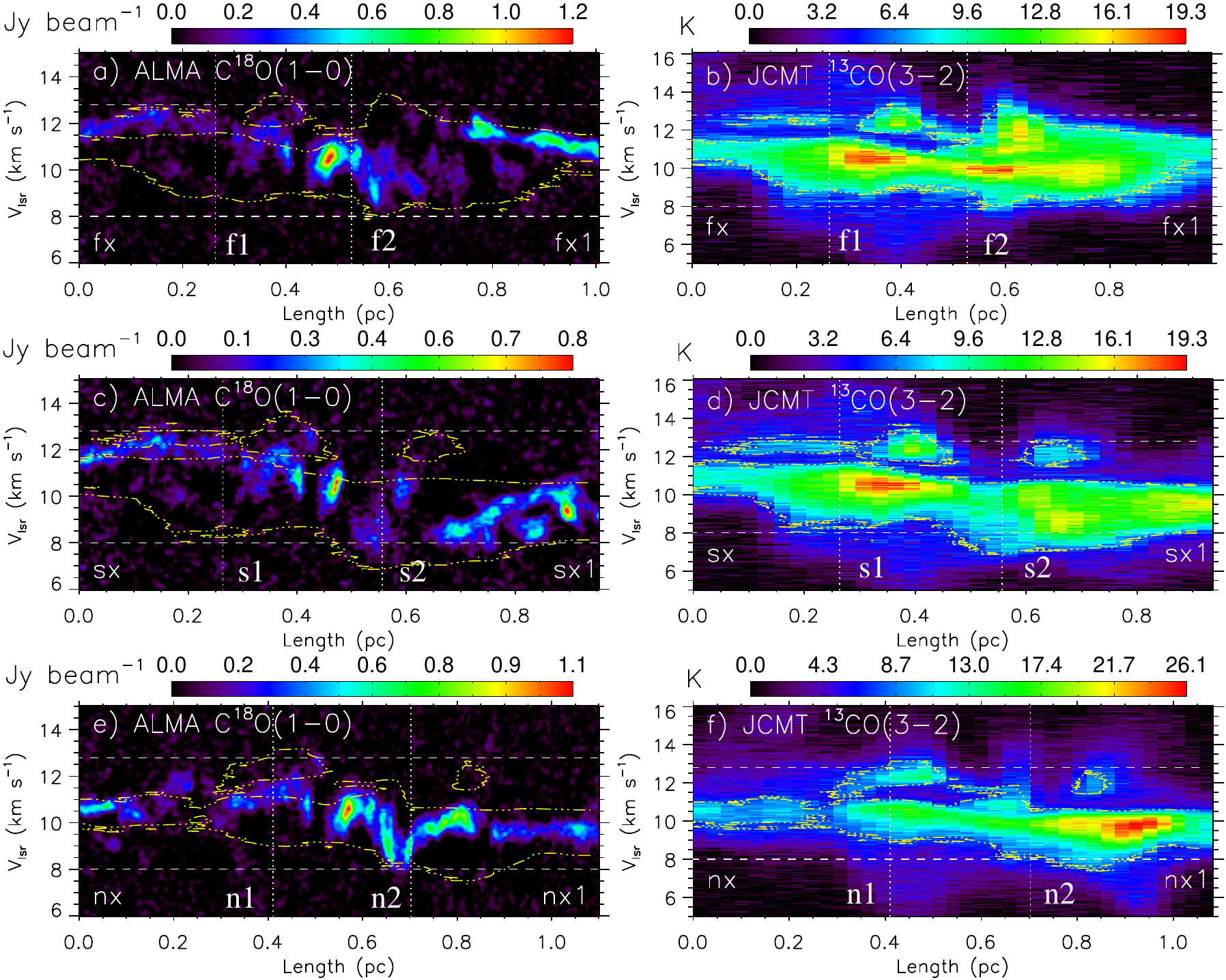}
\caption{a, c, e) PV diagrams of the ALMA C$^{18}$O(1--0) emission extracted along the paths fx--fx1, sx--sx1, and nx--nx1, respectively. b, d, f) Corresponding PV diagrams of the JCMT $^{13}$CO(3--2) emission along the same paths. The path sx--sx1 is indicated in Figure~\ref{fig3}a, while the paths fx--fx1 and nx--nx1 are marked in Figure~\ref{fig3}b. In each panel, vertical dotted lines indicate the locations where the respective paths connect to the central hub/molecular ring; these positions are also labeled in the corresponding PV diagrams (see also Figures~\ref{fig3}c and~\ref{fig3}d). Each PV diagram is extracted along a one-pixel-wide path. The JCMT $^{13}$CO(3--2) emission is overlaid as a dot-dashed contour at 7.5, 6.0, and 8.0 K in panels (a,b), (c,d), and (e,f), respectively.}
\label{fig4}
\end{figure*}

Figure~\ref{fig4} presents the PV diagrams of the ALMA C$^{18}$O(1--0) and JCMT $^{13}$CO(3--2) emissions extracted along the three selected paths (fx--fx1, sx--sx1, and nx--nx1; see Figures~\ref{fig3}a and~\ref{fig3}b). Each path crosses the central hub/molecular-ring region and extends into the surrounding filamentary structures, thereby sampling the velocity field along different spatial directions. The hub connection points, s1--s2, f1--f2, and n1--n2, are marked by vertical dotted lines in the corresponding PV diagrams and labeled accordingly (see also Figures~\ref{fig3}c and~\ref{fig3}d). These provide common spatial references for examining the velocity structures along the filamentary sections (fx--f1, sx--s1, nx--n1, f2--fx1, s2--sx1, and n2--nx1) and across the central hub/molecular-ring segments (f1--f2, s1--s2, and n1--n2). Note that the paths sx--s1 and fx--f1 trace the same prominent filament. 

The PV diagrams reveal systematic velocity structures along the filamentary sections and across the central hub/molecular-ring region (Figure~\ref{fig4}). Along the filamentary sections fx--f1, sx--s1 and nx--n1, which extend from the outer regions toward the central hub, the ALMA C$^{18}$O(1--0) emission shows relatively coherent velocity structures, with the emission predominantly distributed around $\sim$8--13~km~s$^{-1}$. Systematic variations in $V_{\rm LSR}$ are observed along these paths as they approach the central region. The corresponding sections on the opposite side of the hub, f2--fx1, s2--sx1, and n2--nx1, also exhibit velocity-coherent emission, although the dominant velocity and the detailed emission distribution vary along each path. At the ALMA resolution, these structures appear as relatively narrow and localized emission features, with several compact substructures along the individual filamentary sections.

The JCMT $^{13}$CO(3--2) PV diagrams trace these filamentary sections within a broader molecular-gas velocity field. Along the fx--fx1 cut (Figure~\ref{fig4}b), the emission forms an extended velocity structure across both sides of the central region, with systematic changes in the dominant $V_{\rm LSR}$ along the cut. The sx--sx1 cut (Figure~\ref{fig4}d) shows two broad velocity components that extend along the filamentary sections on either side of the central region. Their velocity separation decreases toward the central hub, particularly across the s1--s2 segment. A similar behavior is seen in the nx--nx1 cut (Figure~\ref{fig4}f), where two broad velocity features extend along the nx--n1 and n2--nx1 sections and become progressively closer toward the central region. For comparison with the higher-resolution ALMA data, the JCMT $^{13}$CO(3--2) emission contour is overlaid on the corresponding PV diagrams (Figure~\ref{fig4}). Compared with ALMA, the JCMT emission is spatially broader and smoother, while ALMA resolves narrower and more localized velocity structures. Thus, the six ALMA-resolved filamentary sections are embedded within the more extended molecular-gas velocity field traced by the JCMT observations. 

Across the central hub/molecular-ring region, represented by the f1--f2, s1--s2, and n1--n2 segments, the PV diagrams show more complex velocity structures than those along the individual filamentary sections. Systematic shifts in $V_{\rm LSR}$ are observed across the central hub/molecular-ring region, with distinct velocity features appearing on either side of the central region. As mentioned, the higher angular resolution of ALMA resolves these structures into compact emission features and localized velocity gradients, whereas the JCMT data reveal a more extended and spatially continuous distribution. This enhanced kinematic complexity toward the center indicates that the velocity field becomes increasingly structured within the hub/molecular-ring region.

Taken together, the ALMA and JCMT PV diagrams show consistent velocity structures across their respective spatial scales, with the ALMA-resolved features embedded within the broader molecular-gas velocity field traced by JCMT. Toward the central hub/molecular-ring region, the velocity structure becomes increasingly complex, with the distinct velocity components becoming progressively closer in velocity and increasingly overlapping.
\section{Discussion}
\label{sec:disc}
We combined the JCMT $^{13}$CO(3--2) and ALMA C$^{18}$O(1--0) observations to investigate the kinematics of the Mon~R2 HFS across multiple spatial scales (Sections~\ref{subsec:cloud} and~\ref{subsec:core}). 
The unpublished JCMT $^{13}$CO(3--2) data show the spatial convergence of multiple velocity-coherent filaments toward the central hub, together with their systematic velocity structure and increasing overlap toward the hub (see Figure~\ref{xfig1}). The JCMT $^{13}$CO PPV and PV diagrams provide further evidence for this kinematic organization, exhibiting butterfly/X-shaped velocity structures. Similar morphologies have previously been associated with velocity convergence and gas accumulation in molecular clouds \citep{schneider10,gomez14,smith16,beuther20}. The observed PV morphology does not show the single, continuous velocity gradient that would be expected from simple large-scale rotation. Instead, the velocity field becomes increasingly complex toward the hub, reflecting the superposition of multiple velocity-coherent structures. The PPV diagrams further show intermediate-velocity emission between these structures, indicating a continuous kinematic connection across the converging filamentary network. Together with the observed spatial convergence of the filaments, these features are consistent with hierarchical gas transport toward the hub through multiple converging streams rather than a single coherent inflow.

To further assess whether the observed PV morphology from the JCMT data can arise from gravitationally organized gas motions, we compared the observations with a phenomenological rotating-and-infalling envelope model (Figure~\ref{nnafg3}b in Appendix~\ref{subsec:a2}). The model reproduces the characteristic X-/butterfly-shaped PV morphology and its principal kinematic features. This comparison demonstrates that a combination of rotation and infall may reproduce the observed velocity structure; however, it does not uniquely distinguish between the underlying physical mechanisms responsible for the large-scale gas motions. In particular, the observed morphology can also arise in a dynamically complex environment involving converging flows and hierarchical gas accretion.

Numerical simulations have shown that HFSs can form through large-scale converging flows that generate filamentary networks and funnel gas into dense hubs, thereby linking cloud-scale dynamics with clustered star formation \citep{semadeni07,smith16,nozaki26}. Such flows may be driven by turbulent compression, global gravitational collapse, or the self-gravity of filamentary networks \citep{myers09,Motte+2018,padoan20,semadeni19}. In particular, the recent model of \citet{nozaki26} predicts that HFSs are assembled through converging large-scale gas flows that form radially aligned filaments and channel material toward a central hub. These simulations further suggest that mass accretion is preferentially transported through dense filamentary structures, while the surrounding lower-density gas exhibits comparatively weaker motions, resulting in increasingly complex kinematics toward the hub center. Our observations are broadly consistent with this picture. 

Most recently, \citet{hwang26} demonstrated that inter-filamentary gas can replenish dense filaments, which subsequently transport material toward the Mon~R2 hub. Our $^{13}$CO(3--2) observations reveal extended emission between and around the filamentary structures identified by DisPerSE, suggesting the presence of an extended, lower-density molecular component within and around the filamentary network (see Figure~\ref{xfig1}). Such a component could provide an additional reservoir for replenishing the denser filaments. The nature and dynamical role of this extended component, however, require further investigation. 

In the present study, the JCMT data reveal parsec-scale velocity-coherent structures that are resolved by the ALMA observations into multiple dense, velocity-coherent filaments converging toward the hub. The spatial and velocity continuity between the JCMT- and ALMA-traced structures provides evidence for a kinematic connection across cloud-to-core scales (see Sections~\ref{1subs} and~\ref{2subs} for more details). Furthermore, the presence of intermediate-velocity emission further supports a continuous kinematic connection between the converging filamentary structures and the central hub. The ALMA-resolved filaments therefore appear to be embedded within the broader molecular-gas velocity field traced by the JCMT, rather than representing an isolated kinematic component. This correspondence suggests that the velocity organization of the molecular gas is maintained from the parsec-scale cloud environment down to the dense filamentary structures surrounding the hub. In Section~\ref{2subs}, the molecular line analysis further shows a progressive reduction in velocity separation toward the hub, indicating that the distinct kinematic structures become increasingly blended and overlap within the central region. The correspondence between the broad JCMT velocity components and the ALMA-resolved kinematic domains suggests that the central hub/molecular-ring region represents a transition zone where multiple large-scale velocity structures converge and give rise to a more complex dense-gas velocity field.

Assuming that the observed velocity convergence traces gas motions directed toward the hub, a characteristic flow timescale can be estimated as $t_{\rm flow}=L/\Delta V$, where $L$ is the convergence length scale and $\Delta V$ is the characteristic velocity difference between the converging components. Adopting $L \sim 0.5$ pc and $\Delta V \sim 2.5$ km\,s$^{-1}$ from the JCMT $^{13}$CO PV diagrams gives $t_{\rm flow} \approx 2\times10^{5}$ yr. For the full filament extent ($L\sim1$ pc), the corresponding timescale is $\sim$4 $\times$ $10^{5}$ yr. These values should be regarded as order-of-magnitude kinematic estimates because projection effects may affect both the inferred flow length and velocity. Nevertheless, the resulting timescales are comparable to the estimated age of the central ultra-compact H\,{\sc ii} region ($\sim$10$^{5}$ yr; \citealt{didelon15}), suggesting that the large-scale gas dynamics occur on a timescale relevant to the current episode of star formation in Mon~R2.

If the observed velocity separation is adopted as a characteristic line-of-sight speed of the converging motions, the corresponding Mach number, $\mathcal{M}=\Delta V/c_{\rm s}$, is $\sim12$ for a molecular-gas sound speed of $c_{\rm s}\approx0.2$ km,s$^{-1}$ ($T\sim10$--15 K). This indicates highly supersonic line-of-sight motions, consistent with large-scale converging flows capable of compressing dense gas and transporting material toward the hub. Such supersonic motions are also expected in theoretical models of HFS formation involving large-scale gas flows and gravitationally driven accretion. The inferred Mach number should nevertheless be regarded as an approximate value because the measured velocity separation represents only the line-of-sight component of the gas motions and may not directly correspond to the three-dimensional flow velocity.

Overall, these findings are consistent with a hierarchical kinematic picture involving multiple dynamical processes.
\section{Summary and Conclusion}
\label{sec:sumc}
We investigated the gas kinematics of the Mon~R2 HFS over multiple spatial scales using JCMT $^{13}$CO(3--2) and ALMA C$^{18}$O(1--0) observations to characterize the organization and assembly of molecular gas toward the central hub. The key findings of this study are summarized below:

\begin{enumerate}

\item The JCMT $^{13}$CO(3--2) observations reveal several velocity-coherent filaments with a broad range of position angles that converge toward the central hub of Mon R2, where massive stars and stellar clusters are located.

\item Two dominant molecular velocity regimes, at $\sim$8.8--10.2 and $\sim$10.4--11.8 km s$^{-1}$, are identified on cloud scales. Their velocity separation progressively decreases toward the central hub, indicating increasing kinematic overlap of the large-scale molecular structures.

\item The JCMT $^{13}$CO(3--2) PV and PPV diagrams exhibit a characteristic butterfly/X-shaped morphology. This structure is consistent with molecular gas motions converging toward the central region. Assuming that the observed velocity convergence traces gas motions directed toward the hub, a characteristic flow timescale of $\sim$0.2--0.4 Myr is estimated.

\item To assess whether the observed large-scale PV morphology can arise from gravitationally organized motions, we compared the JCMT data with a phenomenological rotating-and-infalling envelope model convolved to the JCMT angular and spectral resolutions. The model reproduces the principal X-/butterfly-shaped morphology and major kinematic features, demonstrating that rotation and infall may contribute to the observed velocity structure. However, the model does not uniquely distinguish between the different physical mechanisms that may produce the observed kinematics.

\item The ALMA C$^{18}$O(1--0) observations resolve the central region into a network of dense, velocity-coherent filaments and substructures spanning $\sim$8--13 km~s$^{-1}$. Three principal kinematic domains ($\sim$9--10.5, 10.5--11.5, and 12--13 km~s$^{-1}$) converge toward the hub, where their increasing spatial overlap produces a complex velocity field.

\item A comparison of the JCMT $^{13}$CO(3--2) and ALMA C$^{18}$O(1--0) velocity structures provides evidence for a multi-scale kinematic connection between the parsec-scale molecular gas and the dense filamentary network in the central region. This comparison is not intended to establish a one-to-one correspondence between individual velocity components. Rather, the ALMA-resolved dense filaments are embedded within the broader velocity field traced by the JCMT observations, demonstrating how the large-scale velocity structure is organized into multiple dense components at higher angular resolution.

\end{enumerate}

Collectively, the cloud-scale and core-scale gas observations reveal converging and increasingly concentrated gas motions toward the Mon~R2 hub, supporting hierarchical gas assembly through converging flows and filamentary accretion. The findings suggest that filamentary gas accretion may continue during the ongoing evolution of the massive star-forming region. The observed kinematics are nevertheless likely influenced by multiple dynamical processes, including rotation, infall, and potentially other mechanisms. Thus, rather than uniquely establishing a single formation scenario, our results demonstrate a coherent multi-scale kinematic connection between the molecular cloud, filamentary network, and dense gas associated with the Mon R2 hub. 
\section*{Acknowledgments}
We thank the anonymous referee for providing useful comments. The research work at Physical Research Laboratory is funded by the Department of Space, Government of India. 
This paper makes use of the following ALMA data: ADS/JAO.ALMA\#2016.1.01144.S. ALMA is a partnership of ESO (representing its member states), NSF (USA) and NINS (Japan), together with NRC (Canada) , MOST and ASIAA (Taiwan), and KASI (Republic of Korea), in cooperation with the Republic of Chile. The Joint ALMA Observatory is operated by ESO, AUI/NRAO and NAOJ. This NVAS image was produced as part of the NRAO VLA Archive Survey, (c) AUI/NRAO. The James Clerk Maxwell Telescope is operated by the East Asian Observatory on behalf of The National Astronomical Observatory of Japan; Academia Sinica Institute of Astronomy and Astrophysics;  the Korea Astronomy and Space Science Institute; the Operation, Maintenance and Upgrading Fund for Astronomical Telescopes and Facility Instruments, budgeted from the Ministry of Finance (MOF) of China and administrated by the Chinese Academy of Sciences (CAS), as well as the National Key R\&D Program of China (No. 2017YFA0402700). Additional funding support is provided by the Science and Technology Facilities Council of the United Kingdom and participating universities in the United Kingdom and Canada.
%
\appendix
\restartappendixnumbering

\section{JCMT and ALMA maps}
\label{sec:apdx}
\subsection{JCMT $^{13}$CO(3--2) channel maps}
\label{subsec:a1}
\begin{figure*}
\centering
\includegraphics[width=\textwidth]{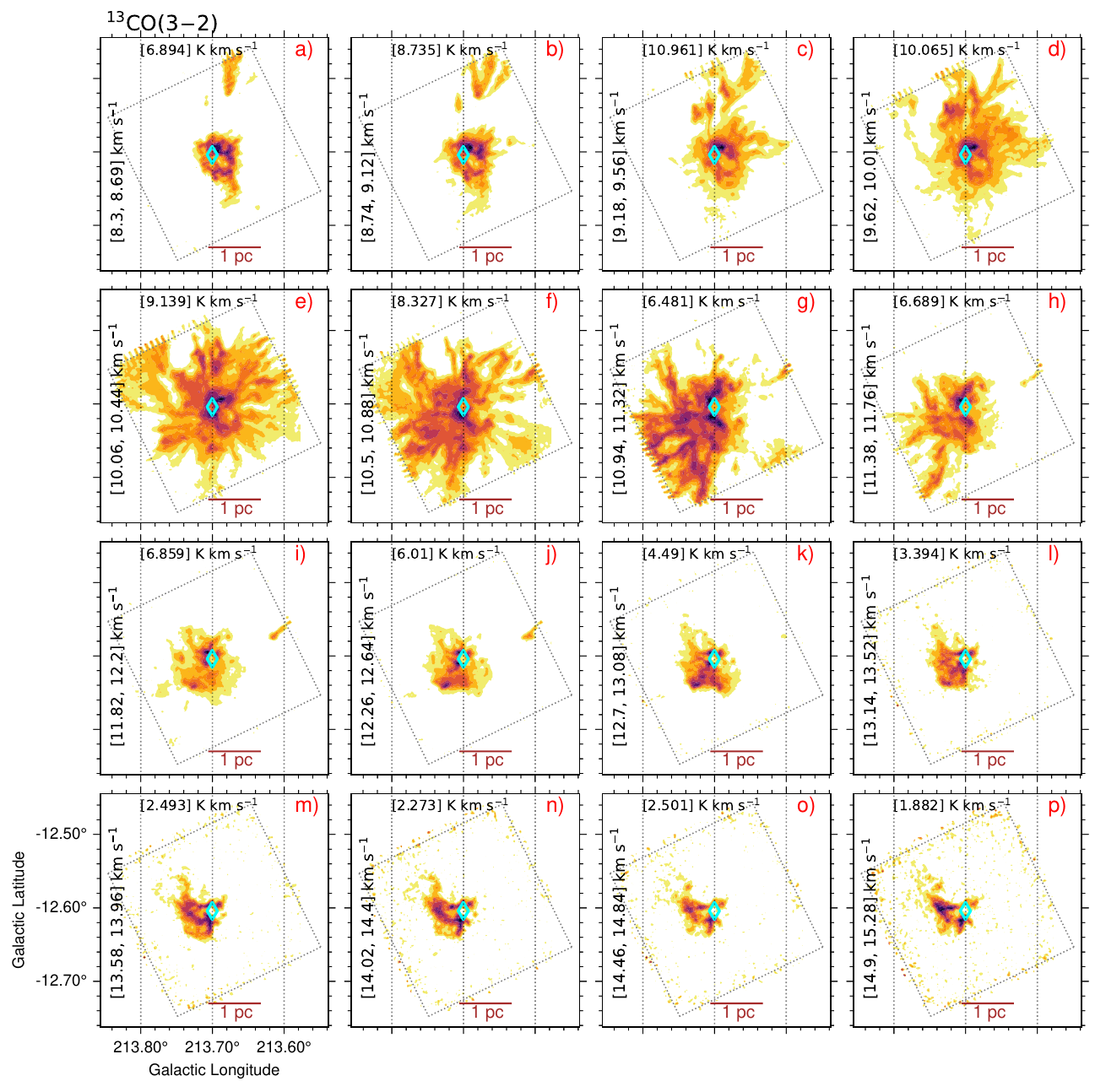}
\caption{Channel maps of the JCMT $^{13}$CO(3--2) emission. The cyan diamond refers to the position of IRS~2. Filled contours in each velocity channel correspond to 
[0.1, 0.2, 0.3, 0.35, 0.4, 0.5, 0.6, 0.7, 0.8, 0.9, 1] $\times$ the peak intensity of that channel, with the peak value listed in each panel. The scale bar indicates 1 pc at the adopted distance of 830 pc.}
\label{afg2}
\end{figure*}

Figure~\ref{afg2} presents the channel maps of the JCMT $^{13}$CO(3--2) emission. The molecular gas displays systematic velocity-dependent morphological variations. 
The low- and high-velocity components exhibit distinct spatial distributions at larger distances from the center, while their emission progressively 
converges and overlaps toward the central HFS. 
\subsection{Position-Velocity Analysis of the $^{13}$CO(3--2): Observations and Model}
\label{subsec:a2}
\begin{figure*}
\centering
\includegraphics[width=\textwidth]{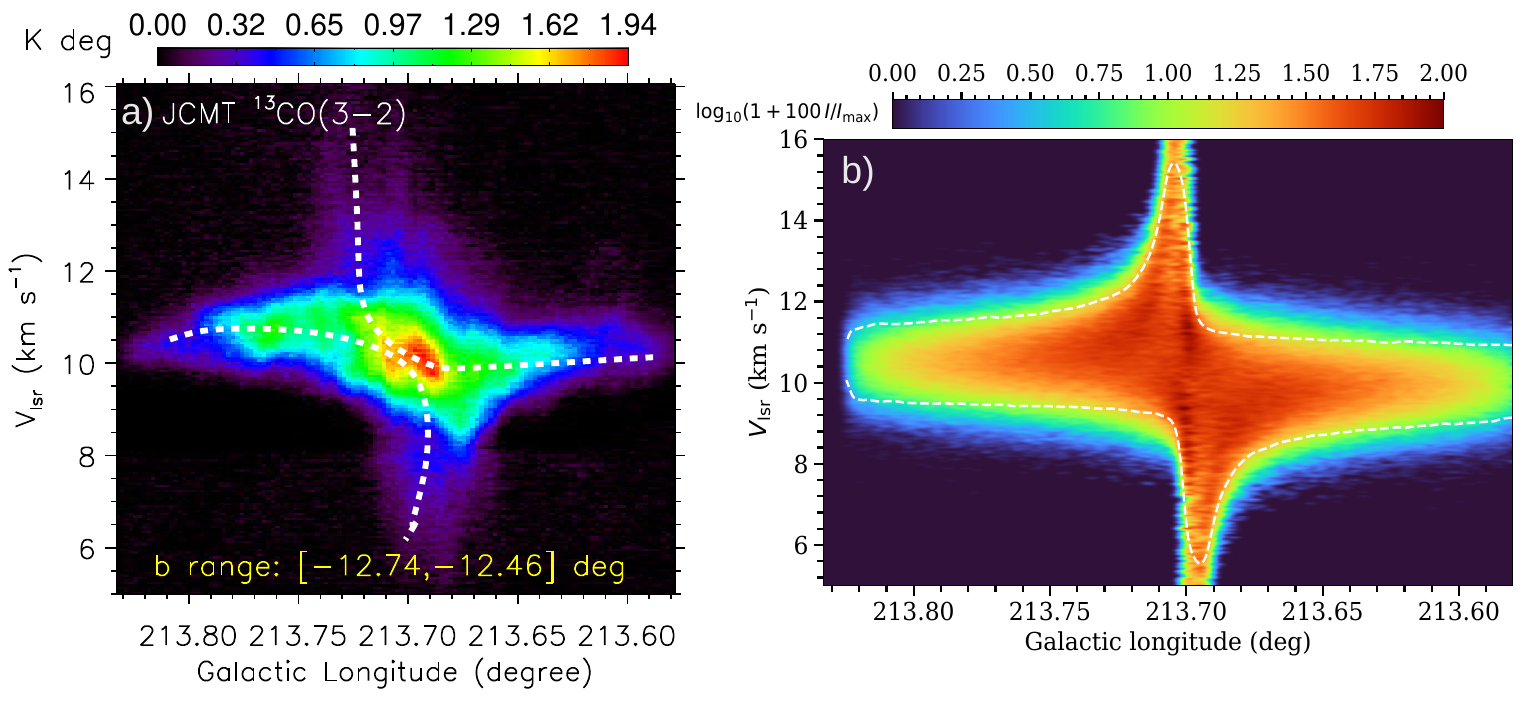}
\caption{a) JCMT $^{13}$CO(3--2) PV diagram toward Mon R2 (see also Figure~\ref{fig2}a). b) Synthetic PV diagram produced from a rotating-infalling envelope model and convolved to the angular and spectral resolutions of the JCMT observations. The white dashed curves represent the intensity-weighted 5th and 95th percentile velocity limits derived directly from the beam- and spectrally convolved model PV diagram. The adopted model parameters are given in Appendix~\ref{subsec:a2}.} 
\label{nnafg3}
\end{figure*}

Figure~\ref{nnafg3}b displays the longitude-velocity diagram of the JCMT $^{13}$CO(3--2) emission, which is the same as shown in Figure~\ref{fig2}a. To further examine whether the observed PV structure can be explained by gravitationally driven motions, we constructed a simple phenomenological rotating-infalling envelope model, motivated by the classical models of rotating gravitational collapse \citep{ulrich76,terebey84} and their applications to massive-star accretion flows \citep{keto10}. We compared its synthetic PV diagram with the JCMT $^{13}$CO(3--2) observations. We adopted a distance of 830 pc and a systemic velocity of $V_{\rm sys}=10.25$ km s$^{-1}$. The model consists of an axisymmetric molecular envelope extending from 0.01 to 1.8 pc, with a density distribution $\rho\propto r^{-1.2}$. The gas is assigned both rotational and radial infall motions. The rotational velocity is described by a constant specific angular momentum, $j=0.5$ km s$^{-1}$ pc, with the rotational speed limited to the local Keplerian velocity, while the radial velocity corresponds to 20\% of the local free-fall velocity. We adopt an enclosed gravitating mass of $1500\,M_\odot$ and an inclination of $60^\circ$. An intrinsic velocity dispersion of 0.6 km s$^{-1}$ is included to account for unresolved motions. 

The model PV diagram was subsequently convolved with the $14^{\prime\prime}$ JCMT beam and the $0.055$ km s$^{-1}$ spectral resolution, thereby allowing a direct comparison with the observed PV diagram. The resulting synthetic PV diagram exhibits a pronounced butterfly-/X-shaped velocity structure, with the velocity extent increasing toward the central position and becoming progressively narrower toward larger angular offsets. This morphology arises naturally from the combined contribution of rotation and radial infall in the projected velocity field. The model therefore reproduces the principal qualitative characteristics of the observed JCMT PV diagram, including the enhanced velocity spread close to the central position and the characteristic narrowing of the emission away from the hub.

For a quantitative comparison, we derived the model velocity envelope directly from the beam- and spectrally-convolved PV diagram rather than from the intrinsic particle velocities. At each longitude, the lower and upper boundaries were defined using the intensity-weighted 5th and 95th percentile velocities of the convolved model emission. As shown by the dashed curves in Figure~\ref{nnafg3}b, the resulting velocity envelope follows the characteristic butterfly-/X-shaped morphology of the synthetic emission. The similarity between the observed velocity structure and the convolved rotating-infalling model indicates that the observed PV morphology is consistent with a combination of rotation and inward radial motions as previously reported by \citet{trevino19} in the Mon R2 hub. Thus, the observed velocity gradient and broad central velocity extent do not require the gas to be interpreted exclusively as rotationally supported material; instead, the PV structure may be naturally produced by an infalling, rotating envelope.
\subsection{ALMA C$^{18}$O(1--0) channel maps}
\label{subsec:a1x}
\begin{figure*}
\centering
\includegraphics[width=\textwidth]{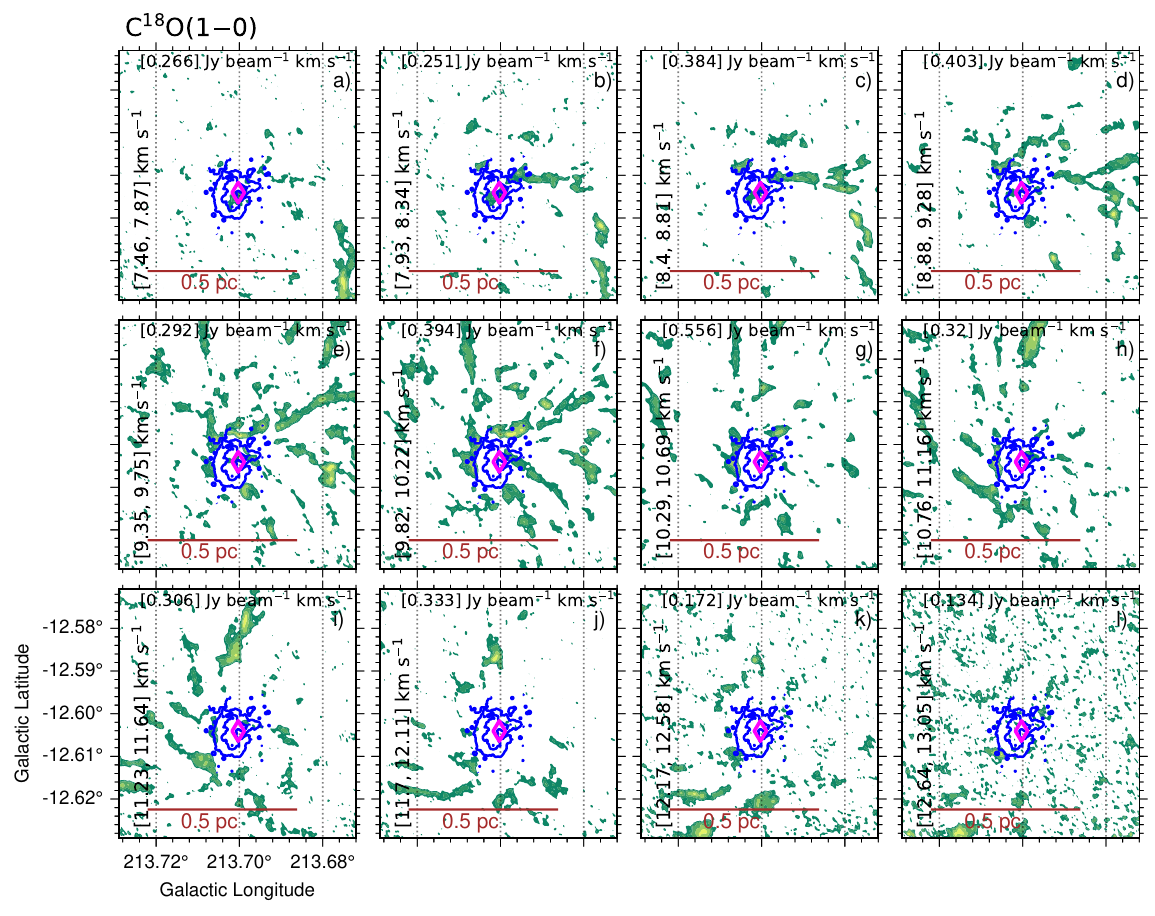}
\caption{Channel maps of the ALMA C$^{18}$O(1--0) emission. The magenta diamond corresponds to the position of IRS~2, and the infrared ring is shown by blue contours. Filled contours in each velocity channel correspond to [0.1, 0.2, 0.3, 0.5, 0.8, 1.0] $\times$ the peak intensity of that channel, with the peak value listed in each panel. The scale bar indicates 0.5 pc at the adopted distance of 830 pc.}
\label{afg3}
\end{figure*}

The ALMA C$^{18}$O(1--0) channel maps are shown in Figure~\ref{afg3}. These maps reveal the detailed kinematic structure of the dense gas within the central hub. 
The emission evolves smoothly across adjacent velocity channels and remains closely associated with the infrared ring and the previously identified infrared sources, highlighting the complex velocity structure of the dense gas in the hub region.
\bibliography{reference}{}
\bibliographystyle{aasjournal}
\end{document}